\documentclass[a4paper,12pt]{article}

\usepackage{amsmath,amssymb}
\usepackage{graphicx}
\usepackage{color}
\usepackage{dcolumn}
\usepackage{bm}
\usepackage[numbers,super,comma,sort&compress]{natbib}
\usepackage{booktabs}

\usepackage{url}
\usepackage{hyperref}
\usepackage{array}
\usepackage{longtable}
\usepackage{diagbox}
\usepackage{multirow}
\usepackage{siunitx}
\usepackage[table]{xcolor}

\newenvironment{wileykeywords}{\textsf{Keywords:}\hspace{\stretch{1}}}{\hspace{\stretch{1}}\rule{1ex}{1ex}}

\usepackage{epstopdf}
\usepackage{subfigure}
\usepackage{indentfirst}
\usepackage{layout} 
\usepackage{authblk}

\title{ML-based Method for Solving the Microkinetic Model of Fischer-Tropsch Synthesis with Varying Catalyst/Reactor Parameters}
\author[1,2]{Taras~Demchuk}
\author[1,3]{Tymofii~Nikolaienko}
\author[4]{Aniruddha~Panda}
\author[4]{Subodh~Madhav~Joshi}
\author[5]{Stanislav~Jaso}
\author[4]{Kaushic~Kalyanaraman}
\affil[1]{SoftServe Inc., 2d Sadova St., 79021 Lviv, Ukraine}
\affil[2]{Institute for Condensed Matter Physics, 1 Svientsitskii St., 79011 Lviv, Ukraine}
\affil[3]{Taras Shevchenko National University of Kyiv, 64/13 Volodymyrska Str., Kyiv 01601, Ukraine}
\affil[4]{Shell India Markets Pvt. Ltd., Mahadeva Kodigehalli, Bengaluru, Karnataka 562149, India}
\affil[5]{Shell Global Solutions International B.V., Grasweg 31, 1031 HW Amsterdam, The Netherlands}
\date{} 

\begin{document}

\maketitle

\begin{abstract}
	This study introduces a physics-informed machine learning framework to accelerate the computation of the microkinetic model of Fischer-Tropsch synthesis. A neural network, trained within the NVIDIA Modulus framework, approximates the fraction of vacant catalytic sites with high accuracy. The combination of implicit differentiation and the Newton-Raphson method enhances derivative calculations, ensuring physical consistency. Computational efficiency improves significantly, with speedups up to $ 10^4 $ times on a GPU. This versatile methodology generalizes across catalysts and reactors, offering a robust tool for chemical engineering applications, including model approximation and catalyst parameter fitting from experimental data.

\end{abstract}

\begin{wileykeywords}
Fischer-Tropsch Synthesis, Mikrokinetic Model, Physics-informed Neural Networks
\end{wileykeywords}

  \makeatletter
  \renewcommand\@biblabel[1]{#1.}
  \makeatother

\bibliographystyle{apsrev}

\renewcommand{\baselinestretch}{1.5}
\normalsize

\clearpage

\section{Introduction}

Over the past two decades, neural networks (NNs) have been widely applied to various scientific and engineering problems, owing to their ability to universally approximate a wide variety of nonlinear functions [\cite{hornikMultilayerFeedforwardNetworks1989}]. Particularly, in chemical engineering, NNs have found extensive use in modeling catalytic reactions and synthesis processes [\cite{pirdashtiArtificialNeuralNetworks2013}, \cite{machadocavalcantiApplicationArtificialNeural2021}, \cite{valeh-e-sheydaApplicationArtificialNeural2010}]. 

One example of such a catalytic process critical to the industry is Fischer-Tropsch synthesis (FTS), which converts syngas (a mixture of $CO$ and $H_2$) into hydrocarbons of varying chain lengths, particularly paraffins and olefins [\cite{mahmoudiReviewFischerTropsch2017}]. This process underpins gas-to-liquid technology, offering an environmentally friendly route to produce clean fuel [\cite{leckelDieselProductionFischerTropsch2009}]. The yield of FTS reactors is highly sensitive to the properties of both the catalyst and the syngas, as well as the overall thermodynamic conditions of the process. 
Developing a robust physical model that accurately describes reactor behavior at multiple scales-integrating microkinetics and macrodynamics enables precise prediction of product distributions and optimization of system parameters to meet specific yield requirements.

Solving the microkinetic model is crucial for multiscale FTS reactor simulations, where, in addition to catalytic surface reactions, the diffusion of reaction components within the reactor is also considered. [\cite{ghouriMultiscaleModelingFixedbed2016}, \cite{luDirectNumericalSimulation2018}]. In such models, the flow of reagents to the catalytic centers is determined by the reaction rate and the rate of product formation value, which can be calculated using the microkinetic model. 
However, such calculations are computationally demanding due to the need for frequent recalculations of reaction rates and product concentrations by solving entire set of microkinetics equations. This challenge motivates the use of NNs to accelerate these computations. Recent work by Patel et al. [\cite{patelAccurateFastFischerTropsch2023, Nikolaienko2024}],
employed physics-informed neural networks (PINNs) [\cite{RAISSI2019686}, \cite{Karniadakis2021}] to solve the main equation of the microkinetic model proposed by Todic et al. [\cite{todicCOinsertionMechanismBased2014}, \cite{todicCorrigendumCOinsertionMechanism2015}]. While this approach significantly accelerated the calculations without sacrificing accuracy, it treated the properties of the catalyst as fixed, requiring time-consuming retraining of the underlying PINN whenever the catalyst material changes. This makes the mentioned approach unusable in data fitting or optimization scenarios, when the multiple catalysts should be analyzed in course of the same simulation.

In this work we generalize the existing approach [\cite{patelAccurateFastFischerTropsch2023}] and propose a new PINN-based model that allows the change of the catalyst properties at the inference stage, eliminating the need in time-consuming retraining. To this end, we derive the modified formulation of the microkinetics model equations, and transform the PINN training process into the space of generalized inputs, suggested by the modified model. The resulting PINN not only enables rapid solution of the initial microkinetics equations, but is also suitable for optimization problems, such as the determination of reaction conditions or other parameters leading to a desired synthesis products distribution. To this end, we will extensively investigate the accuracy of approximating the derivative of the outputs of the proposed PINN model with respect to the system parameters.

The rest of the paper is organized as follows. Section~\ref{sec:mikrokinetic} reviewing the FTS microkinetic model by Todic et al is followed by a general overview of the PINN approach as a method of solving the corresponding type of equations in section \ref{sec:pinn}.
In section~\ref{sec:eq_new}, we adapt these equations for making them agnostic to specific properties of the catalyst and put them into a form suitable for NN training. The description of 
inputs ranges and datasets needed for the proposed PINN training are presented in sections~\ref{sec:parameters} and \ref{sec:datasets}, respectively. 
Section~\ref{sec:PINN} evaluates the performance of the trained PINN. 
The revealed challenges in derivative calculations and solutions using the implicit differentiation approach are discussed in Sections~\ref{sec:fluctuations}, \ref{sec:ID_approach}, and \ref{sec:PINN+ID}.
The computational speedup achieved by the proposed method is analyzed in Section~\ref{sec:time}, followed by final conclusions in Section~\ref{sec:conclusions}.

\section{Methodology}

\subsection{Todic's microkinetic model}
\label{sec:mikrokinetic}

The conversion of syngas into hydrocarbons with varying chain lengths involves multiple reaction steps. The Fischer-Tropsch microkinetics model aims to identify these key steps and describe them through reaction rates and kinetic coefficients. In this study, we employ the Fischer-Tropsch model proposed by Todic et al. [\cite{todicCOinsertionMechanismBased2014}, \cite{todicCorrigendumCOinsertionMechanism2015}], which utilizes the CO-insertion mechanism described in [\cite{Storster2006MicrokineticMO}]. This model divides the reaction into four primary steps: (1) adsorption of the reactants ($ CO $ and $ H_2 $); (2) activation of CO (or the initiation of the hydrocarbon chain); (3) growth of the hydrocarbon chain; and (4) termination of chain growth and desorption (product formation).

In the steady state, the number of catalytic sites occupied by different species remains constant, governed by thermodynamic conditions and catalyst properties.
The microkinetic model is based on the balance between free and occupied catalytic sites. The hydrocarbons production rates with a specific carbon chain length are determined by the fraction of vacant catalytic sites $[S]$ and the chain grows probabilities. 

The balance equation for $[S]$, as defined in [\cite{todicCOinsertionMechanismBased2014}, \cite{todicCorrigendumCOinsertionMechanism2015}], is expressed as follows (for consistency, we maintain the notation used in [\cite{todicCOinsertionMechanismBased2014}] and [\cite{patelAccurateFastFischerTropsch2023}]):
\begin{equation}\label{eq:1}
	[S]=\frac{1}{c_0 + c_S(\alpha_1 + \alpha_1 \alpha_2 + \alpha_1 \alpha_2 \sum_{i=3}^{N_{max}} \prod_{j=3}^{i} \alpha_j)}
\end{equation} 
where $c_0$ and $c_S$ are known analytical functions of thermodynamic and catalytic properties.
Qualitatively, $c_0$ corresponds to the adsorption of $CO$ and $H_2$, and $c_S$ is related to product formation, chain growth, and water formation.
The chain growth probabilities, $\alpha_j$, vary with the carbon number $j$. Due to the specific formation mechanisms of methane and ethylene, $\alpha_1$ and $\alpha_2$ are calculated differently, whereas, for $j \geq 3$, the chain growth probability follows a unified formulation. Detailed physical interpretations and the derivations of each parameter in Eq.~(\ref{eq:1}) can be found in [\cite{patelAccurateFastFischerTropsch2023}].

Solving Eq.~(\ref{eq:1}) provides the fraction of vacant catalytic sites, $[S]$. Once $[S]$ is determined, the probabilities of carbon chain growth, $\alpha_j$, can be calculated, enabling the determination of product formation rates. These rates fully characterize the progression of the Fischer-Tropsch synthesis reaction.

While Eq.~(\ref{eq:1}) can be solved using classical root-finding methods, its nonlinearity imposes substantial computational demands, when multiple instances of the equation are to be solved simultaneously. 
To address this challenge, we adopt the PINN approach, as proposed in [\cite{patelAccurateFastFischerTropsch2023}], to significantly reduce computation time without compromising accuracy.

\subsection{PINN approach details}
\label{sec:pinn}

Unlike conventional data-driven neural network methods, the PINN approach integrates theoretical constraints into the learning process. In this study, the PINN is implemented by formulating the loss function based on the discrepancy between the left-hand side (\textit{l.h.s}) and right-hand side (\textit{r.h.s}) of Eq.~(\ref{eq:1}). Specifically, the workflow is as follows: randomly generated  input parameters (
uniformly distributed within predefined limits) are fed into the NN input layer. The neural network computes outputs corresponding to these inputs. These outputs are used to evaluate the \textit{l.h.s} and \textit{r.h.s} of Eq.~(\ref{eq:1}). The loss function is defined as the average difference between the \textit{l.h.s} and \textit{r.h.s} over a batch of input configurations. This guides the adjustment of the NN's weights during training. The training process continues for a number of steps which was selected large enough to ensure that the neural network accurately approximates the solution. 

Additionally, the PINN approach allows for the inclusion of supplementary constraints, such as boundary conditions or derivative-based losses, in the loss function. This feature is usually used for solving differential equations [\cite{Karniadakis2021, Farea2024}]. However, since this work consider non-differential equation, these advanced capabilities are not utilized in this work.

The training and inference were conducted using NVIDIA Modulus\footnote{Since version 25.03, the package has been renamed to NVIDIA PhysicsNeMo.} [\cite{Modulus_Contributors_NVIDIA_Modulus_An_2023}], a PyTorch-based framework designed specifically for PINN applications. The hyperparameters and architecture of the NN were chosen based on the recommendations in [\cite{patelAccurateFastFischerTropsch2023}], employing four hidden layers with 512 nodes each with fully connected architecture. The GELU [\cite{GELU}] function was used as an activation function while the ADAM [\cite{ADAM}] algorithm was used for optimization. Initially, the learning rate was $ 10^{-3} $ which decreased to $ 10^{-6} $ during the learning process using an exponential scheduler over $ 3\cdot10^{6} $ epochs.

\section{Results and Discussion}
\label{sec:results}

\subsection{Reparametrized equation}\label{sec:eq_new}

In the original work by Patel et al. [\cite{patelAccurateFastFischerTropsch2023}], the variable $[S]$ was modeled as a function of four thermodynamic parameters. 
If it became necessary to include catalyst parameters in their analysis, the total number of variables would rise to 25. (as determined by the balance equation). 
We propose to regroup certain parameters in order to reduce the number of variables (meanings of the symbols introduced here are consistent with those in [\cite{todicCOinsertionMechanismBased2014}, \cite{todicCorrigendumCOinsertionMechanism2015}] and [\cite{patelAccurateFastFischerTropsch2023}]).

According to the $ \alpha_j $ equation three lumped parameters may be introduced:
\begin{align}\label{eq:Snew}
	\tilde{S} = \frac{k_8}{k_3}\frac{1}{K_1 P_{CO}[S]} \\
	\label{eq:A}
	A = \frac{k_8}{k_3}\frac{\sqrt{K_2 P_{H_2}}}{K_1 P_{CO}} \\
	 \label{eq:f}
	 f = e^{-\frac{\Delta E}{k_B T}}
\end{align} 

Dividing both sides of Eq.~(\ref{eq:1}) by $ \frac{k_8}{k_3}\frac{1}{K_1 P_{CO}}$, we obtain:
\begin{equation}\label{eq:new1}
	\frac{k_8}{k_3}\frac{1}{K_1 P_{CO}[S]} = \frac{k_8}{k_3}\frac{1}{K_1 P_{CO}} c_0 + \frac{k_8}{k_3}\frac{1}{K_1 P_{CO}} c_S \alpha_1 \left( 1 + \alpha_2 + \alpha_2 \sum_{i=3}^{N_{max}} \prod_{j=3}^{i} \alpha_j \right).
\end{equation}

Since $\alpha_2$ is the only term involving $k_{8E}$, we introduce an additional parameter:

\begin{align}\label{eq:xi}
	\xi = \frac{k_{8E}}{k_8}
\end{align}

Rewriting $\alpha_2$ and $\alpha_j$ using Eqs.~(\ref{eq:Snew})-(\ref{eq:f}) and (\ref{eq:xi}) gives:

\begin{align}
	\alpha_2 &= \frac{1}{1+A+\xi f^2 \tilde{S}}, \\
	\label{eq:aj}
	\alpha_j &= \frac{1}{1+A+f^j \tilde{S}} \quad (j \geq 3).
\end{align}

From Eq.~(\ref{eq:Snew}), the \textit{l.h.s} of Eq.~(\ref{eq:new1}) simplifies to $\tilde{S}$. To simplify the \textit{r.h.s}, we redefine two terms as follows:

\begin{align}
	\tilde{c}_0 &= \frac{k_8}{k_3}\frac{1}{K_1 P_{CO}} c_0, \\
	\tilde{c}_S &= \frac{k_8}{k_3}\frac{1}{K_1 P_{CO}} c_S \alpha_1.
\end{align}

Substituting these terms into Eq.~(\ref{eq:new1}), the reparametrized equation becomes:

\begin{equation}\label{eq:master_new}
	\tilde{S} = \tilde{c}_0 + \tilde{c}_S \left( 1 + \alpha_2 + \alpha_2 \sum_{i=3}^{N_{max}} \prod_{j=3}^{i} \alpha_j \right).
\end{equation}

This new form expresses $\tilde{S}$ as a nonlinear function of five lumped parameters ($\tilde{c}_0$, $\tilde{c}_S$, $\xi$, $A$, and $f$), which in turn depend on both thermodynamic and catalyst/reactor parameters. The simplified form is well-suited for applying the PINN approach to approximate $\tilde{S}$. Once $\tilde{S}$ is determined using a neural network (NN), $[S]$ can be recovered using Eq.~(\ref{eq:Snew}).

\begin{figure}[h!]
	\begin{centering}
		\includegraphics[width=0.8\linewidth]{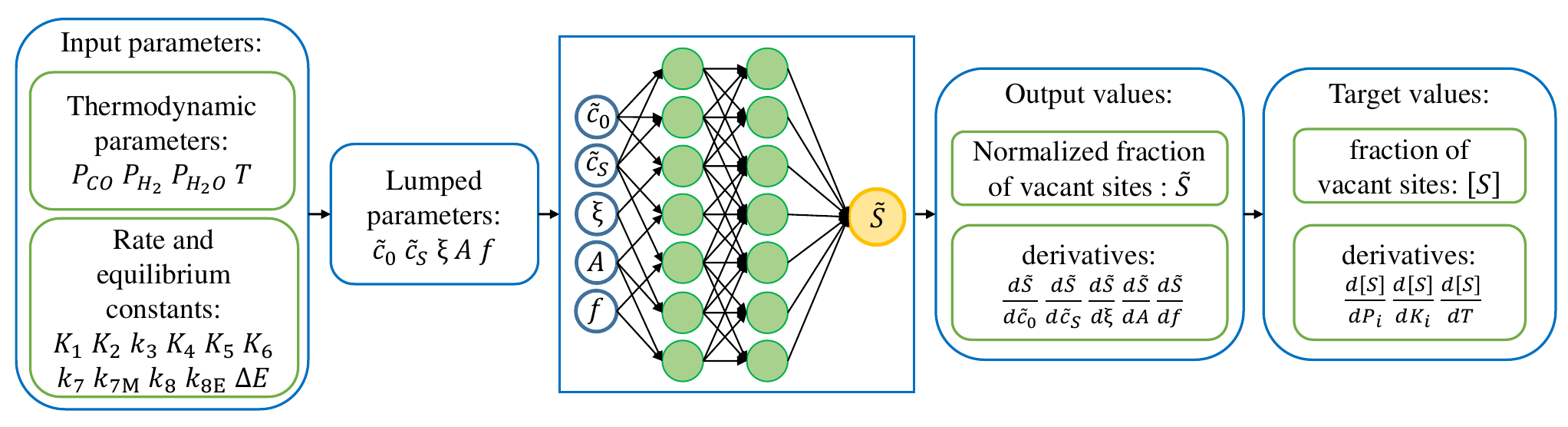}
		\caption{
			Block-scheme of the proposed method
		}
		\label{fig_scheme}
	\end{centering}
\end{figure}

The proposed workflow is illustrated in Fig.~\ref{fig_scheme}. Starting with the 25 input parameters of Todic's microkinetic model, the lumped parameters are calculated, followed by $\tilde{S}$ using the NN, and finally $[S]$ via Eq.~(\ref{eq:Snew}). This approach also facilitates the calculation of derivatives with respect to thermodynamic or catalyst parameters, enabling its use for kinetic parameter fitting or optimizing catalyst properties for desired product distributions.

\subsection{Parameters ranges estimation}\label{sec:parameters}

Training the neural network to approximate $\tilde{S}$ based on the five input parameters defined in Eq.~(\ref{eq:master_new}) requires an understanding of the input parameter ranges. To develop a universal model adaptable to various catalyst and reactor conditions, we estimate the maximum possible boundaries for these inputs based on literature data.

Numerous studies, including [\cite{moazamiComprehensiveStudyKinetics2017}, \cite{ailBiomassLiquidTransportation2016}, \cite{teimouriKineticsSelectivityStudy2021}, \cite{vanderlaanKineticsSelectivityFischer1999}], indicate that Fischer-Tropsch synthesis (FTS) reactors typically operate within a temperature range of approximately 450--650~K. From [\cite{todicCOinsertionMechanismBased2014}, \cite{laanKineticsSelectivityScale1999}], the activation energy parameter $\Delta E$ in Eq.~(\ref{eq:f}) is reported to be approximately 1.0--1.3~kJ/mol. Accounting for a 50\% uncertainty in $\Delta E$ and considering the operational temperature limits, the range for $f$ is estimated to be $(0.5, 1)$. 

The quantity $ \xi $ may be expressed as a fraction of $ C_2 H_4 $ and $ C_3 H_6 $ product formation rates (equations (10) and (11) in [\cite{todicCOinsertionMechanismBased2014}]):
\begin{equation}\label{key}
	\xi = \dfrac{R_{C_2 H_4}}{R_{C_3 H_6}} \alpha_3 f.
\end{equation}
Considering that $ R_{C_3 H_6} > R_{C_2 H_4} $ [\cite{todicCOinsertionMechanismBased2014}, \cite{todicFischerTropschSynthesis2016}] and $ \alpha_3 \in (0,1) $, $\xi$ is constrained to the interval $(0, 1)$. According to the form of equation (\ref{eq:aj}), for infinitely long carbon chains, the chain growth probability $\alpha_\infty$ becomes constant:
\begin{equation}\label{eq:alpha_infinity}
	\alpha_\infty = \frac{1}{1+A+f^\infty \tilde{S}} = \frac{1}{1+A}.
\end{equation}
Industrial considerations and previous studies [\cite{ailBiomassLiquidTransportation2016}, \cite{vervloetFischerTropschReaction2012}, \cite{vanderlaanKineticsSelectivityFischer1999}] suggest that $\alpha_\infty \in (0.2, 1)$. Thus, $A$ is approximately within the range $(0, 4)$. Using Eq.~(\ref{eq:alpha_infinity}) and Eq.~(\ref{eq:aj}), $\tilde{S}$ can be expressed as:
\begin{equation}\label{eq:S_alpha_relationship}
	\alpha_j = \frac{1}{1+A+f^j \tilde{S}} = \frac{1}{\alpha_\infty^{-1} + f^j \tilde{S}} \quad \rightarrow \quad \tilde{S} = \left(\dfrac{1}{\alpha_j} - \dfrac{1}{\alpha_\infty}\right) \dfrac{1}{f^j} \quad \quad j\geq3
\end{equation}
The behavior of $\alpha_j$ as a function of $j$ is qualitatively universal, with $\alpha_j$ increasing for $j \geq 3$ (e.g., Fig.~5 in [\cite{todicCOinsertionMechanismBased2014}] or Fig.~9.12 in [\cite{davisAdvancesFischerTropschSynthesis2010}]). Assuming $\alpha_3$ is at most 10 times smaller than $\alpha_\infty$, and considering the range of $\alpha_\infty$, $\tilde{S}$ is estimated to lie within $(0, 360)$. Consequently, $\tilde{c}_0 \in (0, 360)$ and $\tilde{c}_S \in (0, 360)$ since the multiplier for $\tilde{c}_S$ is always greater than one.

It should be noted that these estimated ranges are based on empirical data and are likely overestimated. For this study, a reduced set of input ranges will be used to simplify the analysis.

\subsection{Explored datasets}\label{sec:datasets}

To evaluate the accuracy of the proposed approach across the entire input hyperspace the dataset named "test dataset" were generated. It consists of $2 \cdot 10^6$ configurations of the parameters $\tilde{c}_0$, $\tilde{c}_S$, $\xi$, $A$, and $f$, generated randomly within the following ranges: $ \tilde{c}_0, \tilde{c}_S, \xi \in (0,1)$, $ A \in (10^{-4},1) $ and $ f \in (0.5,1) $. These ranges are consistent with those used for training all PINN models, as discussed in later sections. The bounds $(0, 1)$ for $\tilde{c}_0$, $\tilde{c}_S$, and $\xi$ were chosen to facilitate normalization, while the range for $A$ begins at a small nonzero value to avoid divergence caused by the asymptotics of $1/A$ in the final term of Eq.~(\ref{eq:master_new}).

For each parameter configuration in the dataset, the value of $\tilde{S}$ was computed using the classical root-finding method implemented via the \textit{fsolve} function from the SciPy package [\cite{Virtanen2020}]. Additionally, the derivatives $ d \tilde{S} / d \tilde{c}_0 $, $ d \tilde{S} / d \tilde{c}_S $, $ d \tilde{S} / d \xi $, $ d \tilde{S} / d A $, and $ d \tilde{S} / d f $ were determined using the implicit differentiation method, which is described in detail in Sec.~\ref{sec:ID_approach}.

\subsection{Performance of the Trained Neural Network}\label{sec:PINN}

The fully connected neural network (FNN) was trained to approximate Eq.~(\ref{eq:master_new}) using the methodology described in Sec.~\ref{sec:pinn}. 

\begin{figure}[h!]
	\begin{centering}
		\includegraphics[width=0.90\linewidth]{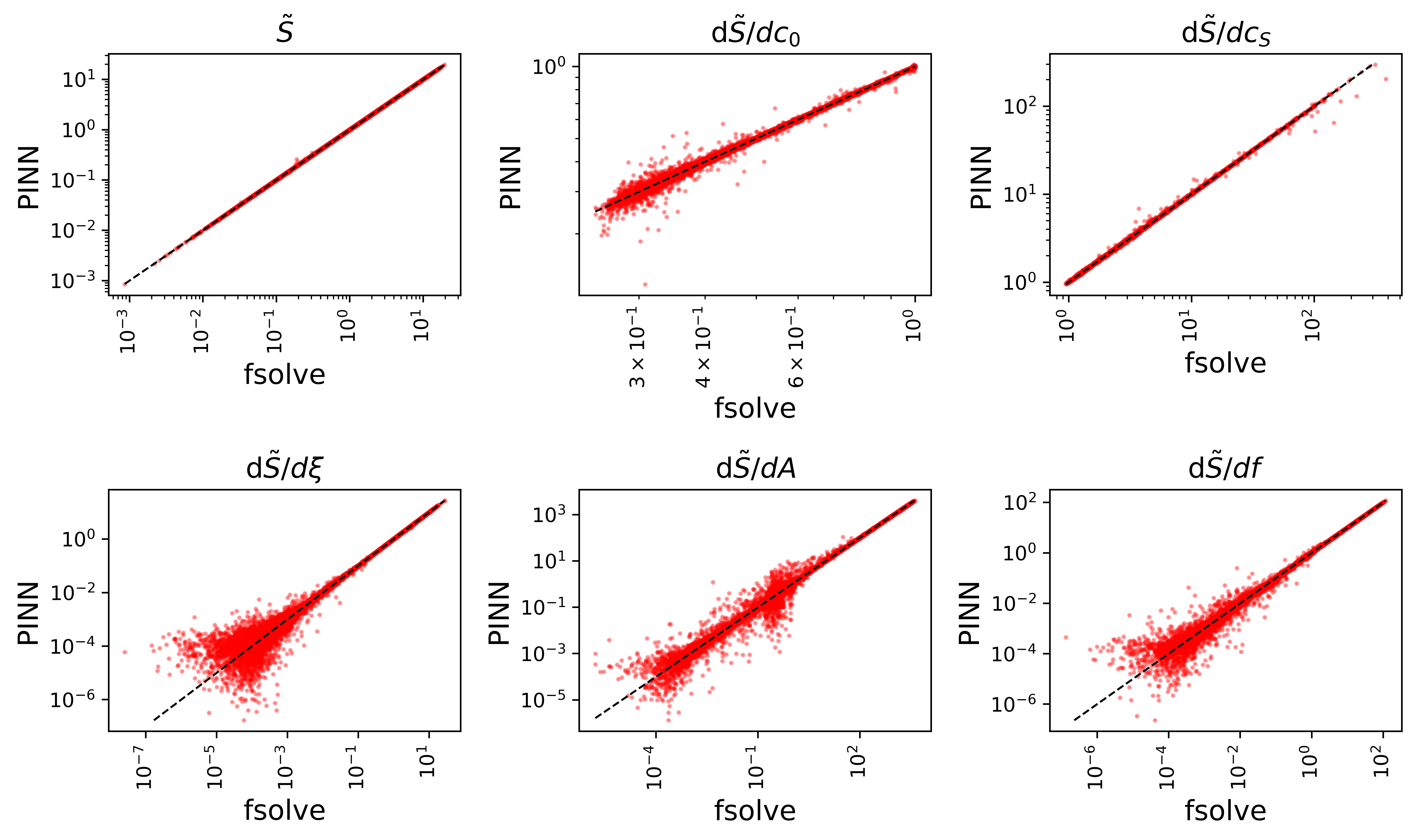}
		\caption{
			PINN-derived vs. \textit{fsolve}-derived values of $ \tilde{S} $ and its derivatives over the Test Dataset.
		}
		\label{fig:33}
	\end{centering}
\end{figure}

To validate the model's performance across the entire input range, $\tilde{S}$ and its derivatives were calculated for the Test Dataset and compared to \textit{fsolve}-derived values. All the NN-derived derivatives were calculated using the automatic differentiation (AD) method [\cite{JMLR:v18:17-468}] as available in NVIDIA Modulus. Fig.~\ref{fig:33} presents these comparisons on a log-log scale, highlighting accuracy across different value magnitudes. The upper left graph compares \textit{fsolve}- and PINN-determined $\tilde{S}$ values, showing that most points lie close to the identity line, indicating high accuracy.

The top center and right graphs contain the same comparison for derivatives $ d \tilde{S} / d \tilde{c}_0 $ and $ d \tilde{S} / d \tilde{c}_S $, respectively. The graph related to $ d \tilde{S} / d \tilde{c}_0 $ values contains a large number of points far from the identity line in the entire range of values, while the graph related to $ d \tilde{S} / d \tilde{c}_S $ contains a small number of such points. This indicates a lower accuracy of the reproduction of $ d \tilde{S} / d \tilde{c}_0 $ and $ d \tilde{S} / d \tilde{c}_S $ values with the help of PINN than it was for $ \tilde{S} $. The lower left, center, and right graphs present the same comparison for derivatives $ d \tilde{S} / d \xi $, $ d \tilde{S} / d A $, and $ d \tilde{S} / d f $, respectively. In all cases, many points deviate from the identity line, particularly at low derivative magnitudes. This suggests that the trained NN struggles with accurately approximating these derivatives.

\begin{table}[h]
	\caption{Average and maximum values of relative errors of trained NN calculated using the Test Dataset}
	\label{tab:33}
	\begin{center}
		\begin{tabular}{|c|c |c |c |c |c |c|}
			\hline
			\rule[-0.5ex]{0pt}{3.0ex} $ \epsilon $ &  $ \tilde{S} $ & $ d \tilde{S} / d \tilde{c}_0 $ &  $ d \tilde{S} / d \tilde{c}_S $ &  $ d \tilde{S} / d \xi $ &  $ d \tilde{S} / d A $ &  $ d \tilde{S} / d f $ \\
			\hline
			
			\rule[-0.5ex]{0pt}{3.0ex} 25\% percentile  & 
			4.50e-06& 5.63e-05& 5.21e-05& 1.92e-04& 1.57e-04& 1.30e-04 \\
			\rule[-0.5ex]{0pt}{3.0ex} 50\% percentile & 
			9.80e-06& 1.26e-04& 1.15e-04& 4.53e-04& 3.75e-04& 3.06e-04 \\
			\rule[-0.5ex]{0pt}{3.0ex} 75\% percentile & 
			1.83e-05& 2.56e-04& 2.19e-04& 1.05e-03& 9.06e-04& 7.01e-04 \\
			\rule[-0.5ex]{0pt}{3.0ex} 99.99\% percentile & 
			2.67e-03& 8.37e-02& 3.27e-02& 1.04e+01& 3.51e+00 & 3.73e+00 \\
			\rule[-0.5ex]{0pt}{3.0ex} mean relative error & 
			1.85e-05& 3.23e-04& 2.27e-04& 1.19e-02& 4.76e-03& 6.09e-03 \\
			\rule[-0.5ex]{0pt}{3.0ex} maximum relative error&  
			3.63e-01& 6.01e-01& 8.40e-01& 2.36e+03& 9.33e+02& 3.34e+03 \\
			\hline
		\end{tabular}
	\end{center}
\end{table}

Table~\ref{tab:33} summarizes the average and maximum relative errors for all calculated quantities over the Test Dataset. Relative errors were calculated according to the following formula:
\begin{equation}
	\epsilon_i = \frac{|X_i^{\mathit{fsolve}}-X_i^{PINN}|}{X_i^{\mathit{fsolve}}}
\end{equation}
While the mean relative error for $\tilde{S}$ is low ($1.85\cdot10^{-5}$), the maximum error is significant ($0.363$).
Nevertheless,  the 99.99th percentile for this sample is 0.00267, indicating the overall suitability of the PINN approach for approximating the Fischer-Tropsch microkinetic model.
However, the NN's accuracy for derivatives is insufficient, limiting its use for 
the synthesis conditions or 
catalyst parameters optimization tasks.

The following subsections explore potential causes of discrepancies in derivative calculations and propose strategies for improvement.

\subsection{Small Fluctuation of Approximated Function}\label{sec:fluctuations}

The use of the GELU activation function, as opposed to the more commonly used ReLU, ensures that derivatives
 of the function learned by the neural network
 are continuous functions. 
However, in certain input value regions, the function $ \tilde{S} (\tilde{c}_0, \tilde{c}_S, \xi, A, f) $ exhibits slow variation, leading to near-zero derivative values. In these regions, the neural network's approximation of the function may experience minor fluctuations around the true value. Consequently, derivatives calculated from the neural network can exhibit disproportionately large magnitudes and deviate significantly from the true values.

\begin{figure}[h]
		\centering
		\subfigure[]{\includegraphics[width=0.45\linewidth]{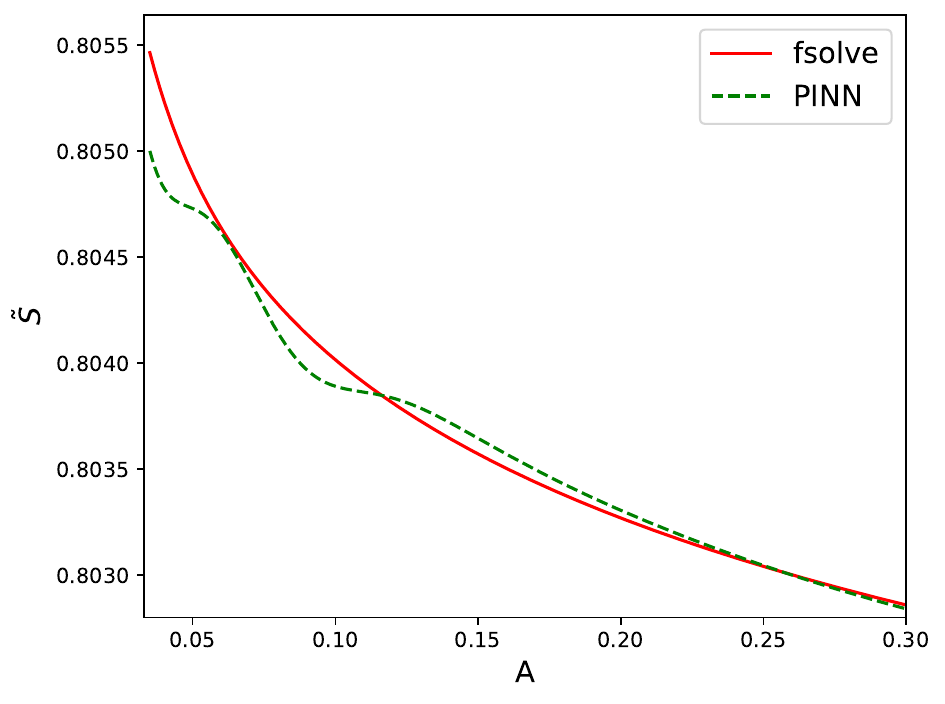}} 
		\subfigure[]{\includegraphics[width=0.45\linewidth]{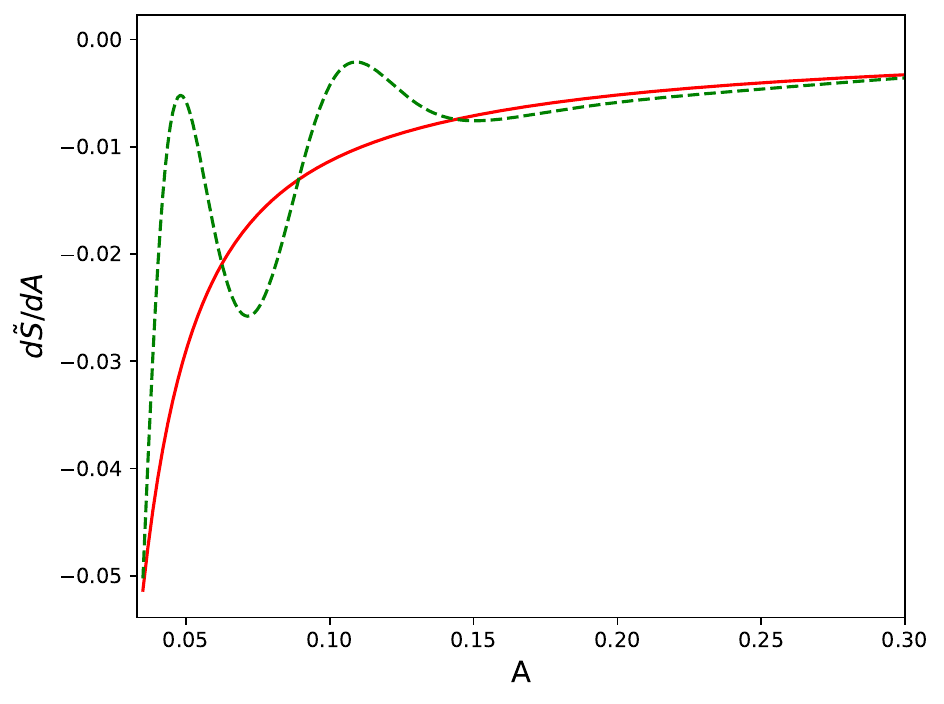}}
		\caption{(a) Dependence of $ \tilde{S} $ on $ A $ calculated by NN and using \textit{fsolve} (b) Dependence of $ d \tilde{S}/ dA $ on $ A $ calculated using the NN and \textit{fsolve}.}
		\label{fig:fluct}
\end{figure}

Figure \ref{fig:fluct} illustrates the dependence of $ \tilde{S} $ and its derivative $ d \tilde{S}/ dA $ (calculated with AD) on the parameter $ A $. The other four input variables were fixed at $ \tilde{c}_0 = 0.8 $, $ \tilde{c}_S = 0.001 $, $ \xi = 0.03 $, and $ f = 0.85 $. These quantities are motivated by Experimental Dataset. 
Figure \ref{fig:fluct}(a) shows that the function derived from the PINN aligns closely with the solver-derived function, with differences becoming apparent only at the fourth decimal place. However, as illustrated in Figure \ref{fig:fluct}(b), the derivatives exhibit significant discrepancies, sometimes differing by an entire order of magnitude. These discrepancies underscore the issue of minor fluctuations in the neural network approximation around the true function, which become amplified when calculating derivatives.

\subsection{Implicit Differentiation Approach}\label{sec:ID_approach}

In order to overcome derivatives calculation issues we propose to use implicit differentiation (ID) method [\cite{Krantz2013}]. This method is useful for finding differentials of functions that are given implicitly and expresses the function's derivative in terms of the values of the function and variables. Using this method will not overcome fluctuations of $ \tilde{S} $, but it will allow us to calculate derivatives based on the accuracy of $ \tilde{S} $ rather than its slope.
This technique is widely used in fields like thermodynamics, reaction kinetics, and systems modeling, where parameters frequently interact in complex, recursive ways [\cite{smith2005introduction}].

A specific method to carry out implicit differentiation in such cases is through the differential approach to implicit functions. This approach involves taking the differential of both sides of an implicit equation, expanding the total differential to account for each variable in the system.

Considering the implicit equation (\ref{eq:master_new}) we introduce a new function $ F(\tilde{S}, \tilde{c}_0 , \tilde{c}_S , \xi , A, f) $ as 
\begin{equation}\label{eq:F}
	F(\tilde{S}, \tilde{c}_0 , \tilde{c}_S , \xi , A, f) = \tilde{S} - \tilde{c}_0 +  \tilde{c}_S  ( 1 + \alpha_2 + \alpha_2 \sum_{i=3}^{N_{max}} \prod_{j=3}^{i} \alpha_j)
\end{equation}
When $\tilde{S}$ is a function of $\tilde{c}_0 , \tilde{c}_S , \xi , A, f) $ that satisfies (\ref{eq:master_new}), turning it into identity in some continuous range of argument values, the newly introduced function $F$ this remains constant ($F \equiv 0$).
The total differential of $ F $ can then be written as:
\begin{equation}\label{eq:dF}
	dF = \frac{\partial F}{\partial \tilde{S}} \, d \tilde{S} + \frac{\partial F}{\partial \tilde{c}_0} d \tilde{c}_0 + \frac{\partial F}{\partial \tilde{c}_S} d \tilde{c}_S + \frac{\partial F}{\partial \xi} d \xi + \frac{\partial F}{\partial A} d A + \frac{\partial F}{\partial f} d f = 0	
\end{equation}
By isolating the terms involving the differential of the $ \tilde{S} $ and the variable of interest $ x $, it is possible to obtain "implicit form" of the partial derivative:
\begin{equation}\label{eq:ID}
	 \frac{d \tilde{S}}{dx} = -\dfrac{\partial F/\partial x}{\partial F/\partial \tilde{S}}
\end{equation} 
This differentiation method provides a systematic approach to computing derivatives for each variable, streamlining the process for analyzing complex systems with interdependent variables [\cite{Marsden2003-nz}, \cite{Rudin1976-nn}].

\subsection{Combination of PINN and ID approach for calculating derivatives}\label{sec:PINN+ID}

Calculating derivatives using the ID method involves two stages: first, for a specific set of variables, the value of $ \tilde{S} $ is calculated, and second, the derivatives are computed using equation (\ref{eq:ID}). In this work, we use the $ \tilde{S} $ values obtained from the trained NN described in the subsection \ref{sec:PINN}. According to Fig.~\ref{fig:33} and Table~\ref{tab:33}, this network reproduces the $ \tilde{S} $ values with high accuracy. On the other hand, derivatives $ \partial F/\partial x $ with $ x \in (tilde{c}_0 , \tilde{c}_S , \xi , A, f) $ and $ \partial F/\partial \tilde{S} $ were calculated using AD.

\begin{figure}[h]
	\centering
	\subfigure[]{\includegraphics[width=0.45\linewidth]{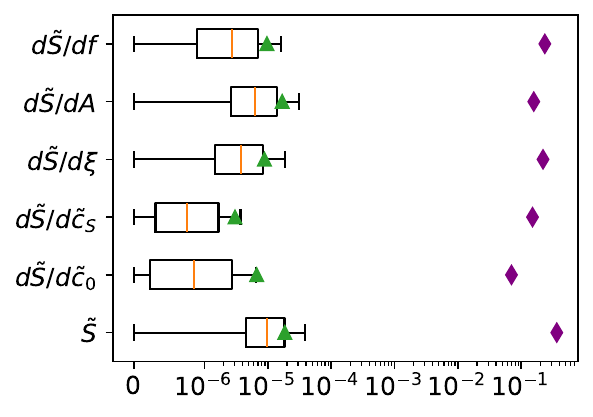}} 
	\subfigure[]{\includegraphics[width=0.45\linewidth]{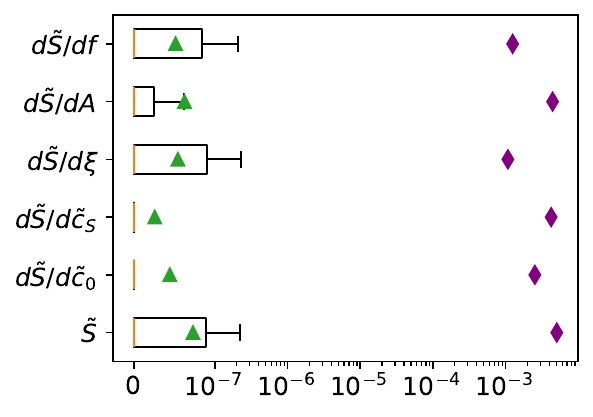}}
	\caption{(a) Boxplots of relative errors in $ \tilde{S} $ and its derivatives calculated over the Test Dataset the using trained NN and ID approach. (b) Boxplots of relative errors in $ \tilde{S} $ and its derivatives calculated over the Test Dataset using the trained NN witn Newton-Raphson correction and ID approach. In both graphs, orange vertical lines corresponds to the median value, green triangels --- to the mean value, and purple diamonds --- to the maximum value.}
	\label{fig:boxplots}
\end{figure}

Figure~\ref{fig:boxplots}(a) presents a boxplot of the relative errors in calculating $ \tilde{S} $ using the trained NN compared to the \textit{fsolve} method, 
as well as boxplots of the relative errors in the derivatives of $ \tilde{S} $ calculated using the ID approach on the Test Dataset.
The yellow vertical line in the boxplots corresponds to the median value, the green triangle indicates the mean value and the diamonds represent the maximum relative errors.
The figure shows that the mean and median values are around $ \sim 10^{-6} $-$ 10^{-5} $ for the $ \tilde{S} $ and its derivatives. However, the maximum relative error in the dataset reaches approximately $ \sim 10^{-1} $.
We consider such accuracy of calculating the $ \tilde{S} $ value and its derivatives to be insufficient for the possible application of NN in optimization problems.

The aforementioned approach for calculating derivatives using the ID method is only valid if identity (\ref{eq:F}) is satisfied. Specifically, the calculated value $ \tilde{S} ( \tilde{c}_0 , \tilde{c}_S , \xi , A, f) $ from the trained NN must ensure that $ F(\tilde{S}, \tilde{c}_0 , \tilde{c}_S , \xi , A, f) $ approaches zero. To improve the accuracy of determining $ \tilde{S} $, we propose using one additional iteration of the Newton-Raphson method [\cite{Kelley2003-ob}]. By evaluating the function $ F $ (eq.~(\ref{eq:F})) for an initial value $ \tilde{S}_0 $, as well as the derivative $ \partial F/\partial \tilde{S}_0 $ at that point, a corrected value $ \tilde{S}_{corr} $ can be calculated, bringing $ F $ closer to zero. The corrected value is calculated as follows:
\begin{equation}\label{eq:corr}
	\tilde{S}_{corr} = \tilde{S}_0 - \dfrac{F(\tilde{S}_0)}{\partial F/\partial \tilde{S} (\tilde{S}_0)}
\end{equation}

The initial values $ \tilde{S}_0 $ are taken from the trained neural network described in the subsection \ref{sec:PINN}, 
while the derivatives are computed using automatic differentiation of $ F(\tilde{S}, \tilde{c}_0 , \tilde{c}_S , \xi , A, f) $ from Eq.~(\ref{eq:F}).The corrected $ \tilde{S}_{corr} $ values are then used to compute derivatives via the ID approach. 

Figure \ref{fig:boxplots}(b) presents a boxplot of the relative errors of $ \tilde{S}_{corr} $ value obtained using eq.~(\ref{eq:corr}) as well as boxplots of its derivatives relative errors calculated using ID approach over the Test Dataset. The figure shows that the maximum relative error for all quantities does not exceed $ 0.005 $. The median value is zero for all quantities, while the average relative error across the test dataset is less than $ 10^{-7} $ for each quantity.

\subsection{Computation Time Estinmation}\label{sec:time}

Using the neural network for finding $ \tilde{S} $ instead of classical root-finding methods is motivated by the necessity for rapid calculations of numerous instances of eq.~(\ref{eq:1}) at the local scale of the catalytic material surface. Moreover, the microkinetic model is calculated simultaneously for each local volume of the reactor for practical use in a multi-scale model [\cite{Wang2001}].

\begin{figure}[h]
	\centering
		\includegraphics[width=0.90\linewidth]{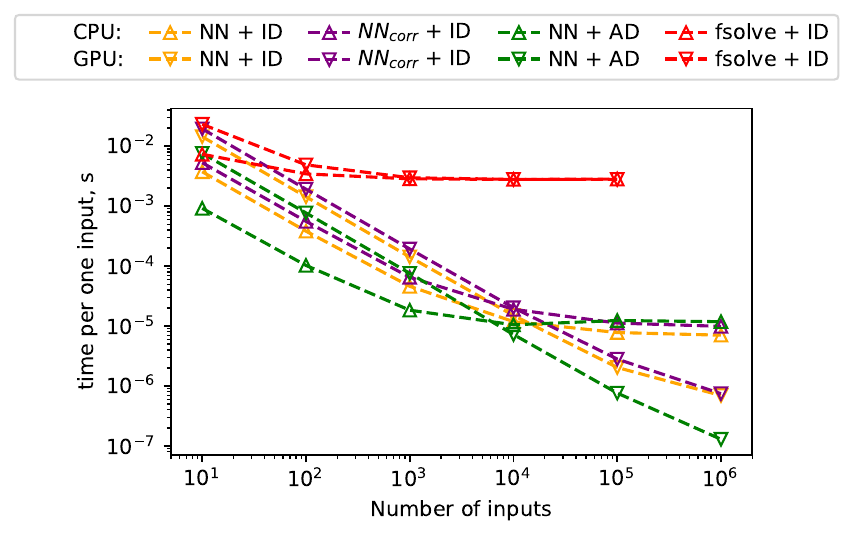}
\caption{
The computational efficiency of the developed PINN in computing $ \tilde{S} $ and its derivatives, shown as the dependence of computation time per input (in seconds) on the number of simultaneously processed inputs for different methods. Different colors represent various methods for these calculations. Upward-facing triangle markers indicate calculations performed on a CPU, while downward-facing triangles correspond to those performed on a GPU (with the \textit{fsolve} function always executed on a CPU).
}
\label{fig:time}
\end{figure}

Figure~\ref{fig:time} presents the computational time required to calculate $ \tilde{S} $ and its derivatives per a single input configuration, as a function of the number of configurations processed simultaneously. 
The lines of different colors correspond to different combinations of methods for calculating $ \tilde{S} $ and its derivatives. 
The abbreviation before the "plus" corresponds to the method for calculating $ \tilde{S} $, in particular, $ \mathtt{NN} $ corresponds to the trained neural network output, $ \mathtt{NN}_{corr} $ is the $ \tilde{S} $ value obtained from the trained network corrected by the Newton method, and $ \mathtt{fsolve} $ is the value of $ \tilde{S} $ obtained from the \textit{fsolve} function. 
The abbreviation after the "plus" corresponds to the method for calculating derivatives of $ \tilde{S} $, in particular, $ \mathtt{AD} $ corresponds to the automatic differentiation method applied to the trained neural network, while $ \mathtt{ID} $ corresponds to the implicit differentiation method. 
For four combinations of the above methods, calculations were performed using both CPU and GPU capabilities. Upward-facing triangle markers indicate calculations performed on a CPU, while downward-facing triangles correspond to those performed on a GPU. It is worth noting that for the line corresponding to the combination "\textit{fsolve} plus ID approach", in calculations using the GPU, only the derivatives were calculated using the GPU using the ID method, while the $ \tilde{S} $ values were calculated using the \textit{fsolve} method on the CPU.

Calculations with the use of CPU capacities allow to evaluation of the speed of various combinations of methods, while GPU-based calculations allow to evaluation of the effectiveness of parallelization for different combinations of methods.

The calculations were performed on a server equipped with an AMD Ryzen 9 5950X processor and an NVIDIA GeForce RTX 4090 GPU, within a Docker container utilizing the NVIDIA Modulus v23.11 image.

The results show that using the \textit{fsolve} function to compute $ \tilde{S} $ and the ID method for derivatives is the slowest approach for large batches. Moreover, when processing $ 10^3 $ or more input configurations simultaneously, the computation speeds become identical for both CPU and GPU implementations.
However, "\textit{fsolve} plus ID approach" demonstrates comparable speed to other methods when calculating 10 or fewer input configurations simultaneously, making it suitable for small batch sizes. 

The use of a NN demonstrates a significant acceleration even for batches of 100 input configurations. Interestingly, for batch sizes smaller than $ 10^4 $, it is more efficient to use the CPU than the GPU for these calculations. On the CPU, the methods "NN plus AD", "NN plus ID", and "corrected NN plus ID" 
exhibit similar computation times that increase with batch size up to $ 10^4 $ inputs, after which computation time remains constant regardless of further batch size growth. 
As a result, employing a NN for calculations achieves up to a 500x speedup compared to the classical method when performed on a CPU.

Using a GPU for NN calculations provides even greater acceleration, with larger input batches reducing per-input computation time. Although this study did not examine batches larger than one million (because of GPU memory limitation), the slope of the curves for all three method combinations suggests a continued decline in per-input computation time as batch sizes exceed one million.

Unlike the case with the CPU, methods "NN plus ID" and "corrected NN plus ID" exhibit comparable speeds, while "NN plus AD" is nearly 10 times faster.

Despite its slightly reduced speed, the "corrected NN plus ID" method achieves significantly higher accuracy in $ \tilde{S} $ and its derivatives, making it the preferred choice. This combination provides a speedup exceeding $ 10^4 $ times compared to the \textit{fsolve} method for batches of one million or more inputs. 

Thus, employing a neural network with corrections and the ID approach is advantageous for both accuracy and computational efficiency in calculating $ \tilde{S} $ and its derivatives. While results may vary depending on the hardware used, the general trends observed in Figure~\ref{fig:time} are expected to remain consistent.

\section{Conclusions}
\label{sec:conclusions}

To summarize, this work describes a method for accelerating the calculation of the microkinetic model of Fischer-Tropsch synthesis using physics-informed machine learning. 
A modified equation for determining the fraction of vacant catalytic sites was obtained by rearranging the thermodynamic and catalytic parameters. This avoided the problem of retraining the neural network for each catalytic material.

A neural network was trained to serve as an approximation function for the given equation within the NVIDIA Modulus framework. By applying the Newton-Raphson method to neural network output the high convergence with the classical root-finding method was achieved. The proposed approach demonstrates the mean relative error of $ 10^{-6} $ magnitude, while the maximum relative error did not exceed $ 10^{-3} $ magnitude.

Challenges associated with derivative computations using automatic differentiation were addressed through the implicit differentiation method, ensuring consistent and accurate results.

This computational scheme delivers substantial speedup compared to classical solvers, achieving up to 500 times faster calculations on a CPU and up to 10,000 times faster on a GPU.

The demonstrated computational efficiency and accuracy position the proposed PINN-based approach as a robust tool for both accelerating the microkinetic modeling process and fitting catalyst parameters to experimental data.

\bibliography{FTS-PINN}

\begin{thebibliography}{35}
\expandafter\ifx\csname natexlab\endcsname\relax\def\natexlab#1{#1}\fi
\expandafter\ifx\csname bibnamefont\endcsname\relax
  \def\bibnamefont#1{#1}\fi
\expandafter\ifx\csname bibfnamefont\endcsname\relax
  \def\bibfnamefont#1{#1}\fi
\expandafter\ifx\csname citenamefont\endcsname\relax
  \def\citenamefont#1{#1}\fi
\expandafter\ifx\csname url\endcsname\relax
  \def\url#1{\texttt{#1}}\fi
\expandafter\ifx\csname urlprefix\endcsname\relax\def\urlprefix{URL }\fi
\providecommand{\bibinfo}[2]{#2}
\providecommand{\eprint}[2][]{\url{#2}}

\bibitem[{\citenamefont{Hornik et~al.}(1989)\citenamefont{Hornik, Stinchcombe,
  and White}}]{hornikMultilayerFeedforwardNetworks1989}
\bibinfo{author}{\bibfnamefont{K.}~\bibnamefont{Hornik}},
  \bibinfo{author}{\bibfnamefont{M.}~\bibnamefont{Stinchcombe}},
  \bibnamefont{and} \bibinfo{author}{\bibfnamefont{H.}~\bibnamefont{White}},
  \bibinfo{journal}{Neural Networks} \textbf{\bibinfo{volume}{2}},
  \bibinfo{pages}{359} (\bibinfo{year}{1989}), ISSN \bibinfo{issn}{08936080}.

\bibitem[{\citenamefont{Pirdashti et~al.}(2013)\citenamefont{Pirdashti,
  Curteanu, Kamangar, Hassim, and
  Khatami}}]{pirdashtiArtificialNeuralNetworks2013}
\bibinfo{author}{\bibfnamefont{M.}~\bibnamefont{Pirdashti}},
  \bibinfo{author}{\bibfnamefont{S.}~\bibnamefont{Curteanu}},
  \bibinfo{author}{\bibfnamefont{M.~H.} \bibnamefont{Kamangar}},
  \bibinfo{author}{\bibfnamefont{M.~H.} \bibnamefont{Hassim}},
  \bibnamefont{and} \bibinfo{author}{\bibfnamefont{M.~A.}
  \bibnamefont{Khatami}}, \bibinfo{journal}{Reviews in Chemical Engineering}
  \textbf{\bibinfo{volume}{29}} (\bibinfo{year}{2013}), ISSN
  \bibinfo{issn}{2191-0235, 0167-8299}.

\bibitem[{\citenamefont{Machado~Cavalcanti
  et~al.}(2021)\citenamefont{Machado~Cavalcanti, Emilia~Kozonoe,
  Andr{\'e}~Pacheco, and Maria De
  Brito~Alves}}]{machadocavalcantiApplicationArtificialNeural2021}
\bibinfo{author}{\bibfnamefont{F.}~\bibnamefont{Machado~Cavalcanti}},
  \bibinfo{author}{\bibfnamefont{C.}~\bibnamefont{Emilia~Kozonoe}},
  \bibinfo{author}{\bibfnamefont{K.}~\bibnamefont{Andr{\'e}~Pacheco}},
  \bibnamefont{and} \bibinfo{author}{\bibfnamefont{R.}~\bibnamefont{Maria De
  Brito~Alves}}, in \emph{\bibinfo{booktitle}{Deep {{Learning Applications}}}},
  edited by \bibinfo{editor}{\bibfnamefont{P.}~\bibnamefont{Luigi~Mazzeo}}
  \bibnamefont{and} \bibinfo{editor}{\bibfnamefont{P.}~\bibnamefont{Spagnolo}}
  (\bibinfo{publisher}{IntechOpen}, \bibinfo{year}{2021}), ISBN
  \bibinfo{isbn}{978-1-83962-374-5 978-1-83962-375-2}.

\bibitem[{\citenamefont{{Valeh-e-Sheyda}
  et~al.}(2010)\citenamefont{{Valeh-e-Sheyda}, Yaripour, Moradi, and
  Saber}}]{valeh-e-sheydaApplicationArtificialNeural2010}
\bibinfo{author}{\bibfnamefont{P.}~\bibnamefont{{Valeh-e-Sheyda}}},
  \bibinfo{author}{\bibfnamefont{F.}~\bibnamefont{Yaripour}},
  \bibinfo{author}{\bibfnamefont{G.}~\bibnamefont{Moradi}}, \bibnamefont{and}
  \bibinfo{author}{\bibfnamefont{M.}~\bibnamefont{Saber}},
  \bibinfo{journal}{Industrial \& Engineering Chemistry Research}
  \textbf{\bibinfo{volume}{49}}, \bibinfo{pages}{4620} (\bibinfo{year}{2010}),
  ISSN \bibinfo{issn}{0888-5885, 1520-5045}.

\bibitem[{\citenamefont{Mahmoudi et~al.}(2017)\citenamefont{Mahmoudi, Mahmoudi,
  Doustdar, Jahangiri, Tsolakis, Gu, and
  LechWyszynski}}]{mahmoudiReviewFischerTropsch2017}
\bibinfo{author}{\bibfnamefont{H.}~\bibnamefont{Mahmoudi}},
  \bibinfo{author}{\bibfnamefont{M.}~\bibnamefont{Mahmoudi}},
  \bibinfo{author}{\bibfnamefont{O.}~\bibnamefont{Doustdar}},
  \bibinfo{author}{\bibfnamefont{H.}~\bibnamefont{Jahangiri}},
  \bibinfo{author}{\bibfnamefont{A.}~\bibnamefont{Tsolakis}},
  \bibinfo{author}{\bibfnamefont{S.}~\bibnamefont{Gu}}, \bibnamefont{and}
  \bibinfo{author}{\bibfnamefont{M.}~\bibnamefont{LechWyszynski}},
  \bibinfo{journal}{Biofuels Engineering} \textbf{\bibinfo{volume}{2}},
  \bibinfo{pages}{11} (\bibinfo{year}{2017}), ISSN \bibinfo{issn}{2084-7181}.

\bibitem[{\citenamefont{Leckel}(2009)}]{leckelDieselProductionFischerTropsch2009}
\bibinfo{author}{\bibfnamefont{D.}~\bibnamefont{Leckel}},
  \bibinfo{journal}{Energy \& Fuels} \textbf{\bibinfo{volume}{23}},
  \bibinfo{pages}{2342} (\bibinfo{year}{2009}), ISSN \bibinfo{issn}{0887-0624,
  1520-5029}.

\bibitem[{\citenamefont{Ghouri et~al.}(2016)\citenamefont{Ghouri, Afzal,
  Hussain, Blank, Bukur, and Elbashir}}]{ghouriMultiscaleModelingFixedbed2016}
\bibinfo{author}{\bibfnamefont{M.~M.} \bibnamefont{Ghouri}},
  \bibinfo{author}{\bibfnamefont{S.}~\bibnamefont{Afzal}},
  \bibinfo{author}{\bibfnamefont{R.}~\bibnamefont{Hussain}},
  \bibinfo{author}{\bibfnamefont{J.}~\bibnamefont{Blank}},
  \bibinfo{author}{\bibfnamefont{D.~B.} \bibnamefont{Bukur}}, \bibnamefont{and}
  \bibinfo{author}{\bibfnamefont{N.~O.} \bibnamefont{Elbashir}},
  \bibinfo{journal}{Computers \& Chemical Engineering}
  \textbf{\bibinfo{volume}{91}}, \bibinfo{pages}{38} (\bibinfo{year}{2016}),
  ISSN \bibinfo{issn}{00981354}.

\bibitem[{\citenamefont{Lu et~al.}(2018)\citenamefont{Lu, Das, Peters, and
  Kuipers}}]{luDirectNumericalSimulation2018}
\bibinfo{author}{\bibfnamefont{J.}~\bibnamefont{Lu}},
  \bibinfo{author}{\bibfnamefont{S.}~\bibnamefont{Das}},
  \bibinfo{author}{\bibfnamefont{E.}~\bibnamefont{Peters}}, \bibnamefont{and}
  \bibinfo{author}{\bibfnamefont{J.}~\bibnamefont{Kuipers}},
  \bibinfo{journal}{Chemical Engineering Science}
  \textbf{\bibinfo{volume}{176}}, \bibinfo{pages}{1} (\bibinfo{year}{2018}),
  ISSN \bibinfo{issn}{00092509}.

\bibitem[{\citenamefont{Patel et~al.}(2023)\citenamefont{Patel, Panda,
  Nikolaienko, Jaso, Lopez, and
  Kalyanaraman}}]{patelAccurateFastFischerTropsch2023}
\bibinfo{author}{\bibfnamefont{H.}~\bibnamefont{Patel}},
  \bibinfo{author}{\bibfnamefont{A.}~\bibnamefont{Panda}},
  \bibinfo{author}{\bibfnamefont{T.}~\bibnamefont{Nikolaienko}},
  \bibinfo{author}{\bibfnamefont{S.}~\bibnamefont{Jaso}},
  \bibinfo{author}{\bibfnamefont{A.}~\bibnamefont{Lopez}}, \bibnamefont{and}
  \bibinfo{author}{\bibfnamefont{K.}~\bibnamefont{Kalyanaraman}},
  \emph{\bibinfo{title}{Accurate and {{Fast Fischer-Tropsch Reaction
  Microkinetics}} using {{PINNs}}}} (\bibinfo{year}{2023}),
  \eprint{2311.10456}.

\bibitem[{\citenamefont{Nikolaienko et~al.}(2024)\citenamefont{Nikolaienko,
  Patel, Panda, Joshi, Jaso, and Kalyanaraman}}]{Nikolaienko2024}
\bibinfo{author}{\bibfnamefont{T.}~\bibnamefont{Nikolaienko}},
  \bibinfo{author}{\bibfnamefont{H.}~\bibnamefont{Patel}},
  \bibinfo{author}{\bibfnamefont{A.}~\bibnamefont{Panda}},
  \bibinfo{author}{\bibfnamefont{S.~M.} \bibnamefont{Joshi}},
  \bibinfo{author}{\bibfnamefont{S.}~\bibnamefont{Jaso}}, \bibnamefont{and}
  \bibinfo{author}{\bibfnamefont{K.}~\bibnamefont{Kalyanaraman}},
  \emph{\bibinfo{title}{Physics-informed neural networks need a physicist to be
  accurate: the case of mass and heat transport in fischer-tropsch catalyst
  particles}} (\bibinfo{year}{2024}),
  \urlprefix\url{https://arxiv.org/abs/2411.10048}.

\bibitem[{\citenamefont{Raissi et~al.}(2019)\citenamefont{Raissi, Perdikaris,
  and Karniadakis}}]{RAISSI2019686}
\bibinfo{author}{\bibfnamefont{M.}~\bibnamefont{Raissi}},
  \bibinfo{author}{\bibfnamefont{P.}~\bibnamefont{Perdikaris}},
  \bibnamefont{and}
  \bibinfo{author}{\bibfnamefont{G.}~\bibnamefont{Karniadakis}},
  \bibinfo{journal}{Journal of Computational Physics}
  \textbf{\bibinfo{volume}{378}}, \bibinfo{pages}{686} (\bibinfo{year}{2019}),
  ISSN \bibinfo{issn}{0021-9991},
  \urlprefix\url{https://www.sciencedirect.com/science/article/pii/S0021999118307125}.

\bibitem[{\citenamefont{Karniadakis et~al.}(2021)\citenamefont{Karniadakis,
  Kevrekidis, Lu, Perdikaris, Wang, and Yang}}]{Karniadakis2021}
\bibinfo{author}{\bibfnamefont{G.~E.} \bibnamefont{Karniadakis}},
  \bibinfo{author}{\bibfnamefont{I.~G.} \bibnamefont{Kevrekidis}},
  \bibinfo{author}{\bibfnamefont{L.}~\bibnamefont{Lu}},
  \bibinfo{author}{\bibfnamefont{P.}~\bibnamefont{Perdikaris}},
  \bibinfo{author}{\bibfnamefont{S.}~\bibnamefont{Wang}}, \bibnamefont{and}
  \bibinfo{author}{\bibfnamefont{L.}~\bibnamefont{Yang}},
  \bibinfo{journal}{Nature Reviews Physics} \textbf{\bibinfo{volume}{3}},
  \bibinfo{pages}{422–440} (\bibinfo{year}{2021}), ISSN
  \bibinfo{issn}{2522-5820},
  \urlprefix\url{http://dx.doi.org/10.1038/s42254-021-00314-5}.

\bibitem[{\citenamefont{Todic et~al.}(2014)\citenamefont{Todic, Ma, Jacobs,
  Davis, and Bukur}}]{todicCOinsertionMechanismBased2014}
\bibinfo{author}{\bibfnamefont{B.}~\bibnamefont{Todic}},
  \bibinfo{author}{\bibfnamefont{W.}~\bibnamefont{Ma}},
  \bibinfo{author}{\bibfnamefont{G.}~\bibnamefont{Jacobs}},
  \bibinfo{author}{\bibfnamefont{B.~H.} \bibnamefont{Davis}}, \bibnamefont{and}
  \bibinfo{author}{\bibfnamefont{D.~B.} \bibnamefont{Bukur}},
  \bibinfo{journal}{Catalysis Today} \textbf{\bibinfo{volume}{228}},
  \bibinfo{pages}{32} (\bibinfo{year}{2014}), ISSN \bibinfo{issn}{09205861}.

\bibitem[{\citenamefont{Todic et~al.}(2015)\citenamefont{Todic, Ma, Jacobs,
  Davis, and Bukur}}]{todicCorrigendumCOinsertionMechanism2015}
\bibinfo{author}{\bibfnamefont{B.}~\bibnamefont{Todic}},
  \bibinfo{author}{\bibfnamefont{W.}~\bibnamefont{Ma}},
  \bibinfo{author}{\bibfnamefont{G.}~\bibnamefont{Jacobs}},
  \bibinfo{author}{\bibfnamefont{B.~H.} \bibnamefont{Davis}}, \bibnamefont{and}
  \bibinfo{author}{\bibfnamefont{D.~B.} \bibnamefont{Bukur}},
  \bibinfo{journal}{Catalysis Today} \textbf{\bibinfo{volume}{242}},
  \bibinfo{pages}{386} (\bibinfo{year}{2015}), ISSN \bibinfo{issn}{09205861}.

\bibitem[{\citenamefont{Stors{\ae}ter et~al.}(2006)\citenamefont{Stors{\ae}ter,
  Chen, and Holmen}}]{Storster2006MicrokineticMO}
\bibinfo{author}{\bibfnamefont{S.}~\bibnamefont{Stors{\ae}ter}},
  \bibinfo{author}{\bibfnamefont{D.}~\bibnamefont{Chen}}, \bibnamefont{and}
  \bibinfo{author}{\bibfnamefont{A.}~\bibnamefont{Holmen}},
  \bibinfo{journal}{Surface Science} \textbf{\bibinfo{volume}{600}},
  \bibinfo{pages}{2051} (\bibinfo{year}{2006}),
  \urlprefix\url{https://api.semanticscholar.org/CorpusID:93281206}.

\bibitem[{\citenamefont{Farea et~al.}(2024)\citenamefont{Farea, Yli-Harja, and
  Emmert-Streib}}]{Farea2024}
\bibinfo{author}{\bibfnamefont{A.}~\bibnamefont{Farea}},
  \bibinfo{author}{\bibfnamefont{O.}~\bibnamefont{Yli-Harja}},
  \bibnamefont{and}
  \bibinfo{author}{\bibfnamefont{F.}~\bibnamefont{Emmert-Streib}},
  \bibinfo{journal}{AI} \textbf{\bibinfo{volume}{5}},
  \bibinfo{pages}{1534–1557} (\bibinfo{year}{2024}), ISSN
  \bibinfo{issn}{2673-2688},
  \urlprefix\url{http://dx.doi.org/10.3390/ai5030074}.

\bibitem[{\citenamefont{{Modulus
  Contributors}}(2023)}]{Modulus_Contributors_NVIDIA_Modulus_An_2023}
\bibinfo{author}{\bibnamefont{{Modulus Contributors}}},
  \emph{\bibinfo{title}{{NVIDIA Modulus: An open-source framework for
  physics-based deep learning in science and engineering}}}
  (\bibinfo{year}{2023}), \urlprefix\url{https://github.com/NVIDIA/modulus}.

\bibitem[{\citenamefont{Hendrycks and Gimpel}(2016)}]{GELU}
\bibinfo{author}{\bibfnamefont{D.}~\bibnamefont{Hendrycks}} \bibnamefont{and}
  \bibinfo{author}{\bibfnamefont{K.}~\bibnamefont{Gimpel}},
  \emph{\bibinfo{title}{Gaussian error linear units (gelus)}}
  (\bibinfo{year}{2016}), \urlprefix\url{https://arxiv.org/abs/1606.08415}.

\bibitem[{\citenamefont{Kingma and Ba}(2014)}]{ADAM}
\bibinfo{author}{\bibfnamefont{D.~P.} \bibnamefont{Kingma}} \bibnamefont{and}
  \bibinfo{author}{\bibfnamefont{J.}~\bibnamefont{Ba}},
  \emph{\bibinfo{title}{Adam: A method for stochastic optimization}}
  (\bibinfo{year}{2014}), \urlprefix\url{https://arxiv.org/abs/1412.6980}.

\bibitem[{\citenamefont{Moazami et~al.}(2017)\citenamefont{Moazami, Wyszynski,
  Rahbar, Tsolakis, and Mahmoudi}}]{moazamiComprehensiveStudyKinetics2017}
\bibinfo{author}{\bibfnamefont{N.}~\bibnamefont{Moazami}},
  \bibinfo{author}{\bibfnamefont{M.~L.} \bibnamefont{Wyszynski}},
  \bibinfo{author}{\bibfnamefont{K.}~\bibnamefont{Rahbar}},
  \bibinfo{author}{\bibfnamefont{A.}~\bibnamefont{Tsolakis}}, \bibnamefont{and}
  \bibinfo{author}{\bibfnamefont{H.}~\bibnamefont{Mahmoudi}},
  \bibinfo{journal}{Chemical Engineering Science}
  \textbf{\bibinfo{volume}{171}}, \bibinfo{pages}{32} (\bibinfo{year}{2017}),
  ISSN \bibinfo{issn}{00092509}.

\bibitem[{\citenamefont{Ail and
  Dasappa}(2016)}]{ailBiomassLiquidTransportation2016}
\bibinfo{author}{\bibfnamefont{S.~S.} \bibnamefont{Ail}} \bibnamefont{and}
  \bibinfo{author}{\bibfnamefont{S.}~\bibnamefont{Dasappa}},
  \bibinfo{journal}{Renewable and Sustainable Energy Reviews}
  \textbf{\bibinfo{volume}{58}}, \bibinfo{pages}{267} (\bibinfo{year}{2016}),
  ISSN \bibinfo{issn}{13640321}.

\bibitem[{\citenamefont{Teimouri et~al.}(2021)\citenamefont{Teimouri,
  Abatzoglou, and Dalai}}]{teimouriKineticsSelectivityStudy2021}
\bibinfo{author}{\bibfnamefont{Z.}~\bibnamefont{Teimouri}},
  \bibinfo{author}{\bibfnamefont{N.}~\bibnamefont{Abatzoglou}},
  \bibnamefont{and} \bibinfo{author}{\bibfnamefont{A.~K.} \bibnamefont{Dalai}},
  \bibinfo{journal}{Catalysts} \textbf{\bibinfo{volume}{11}},
  \bibinfo{pages}{330} (\bibinfo{year}{2021}), ISSN \bibinfo{issn}{2073-4344}.

\bibitem[{\citenamefont{Van Der~Laan and
  Beenackers}(1999)}]{vanderlaanKineticsSelectivityFischer1999}
\bibinfo{author}{\bibfnamefont{G.~P.} \bibnamefont{Van Der~Laan}}
  \bibnamefont{and} \bibinfo{author}{\bibfnamefont{A.~A. C.~M.}
  \bibnamefont{Beenackers}}, \bibinfo{journal}{Catalysis Reviews}
  \textbf{\bibinfo{volume}{41}}, \bibinfo{pages}{255} (\bibinfo{year}{1999}),
  ISSN \bibinfo{issn}{0161-4940, 1520-5703}.

\bibitem[{\citenamefont{van~der Laan}(1999)}]{laanKineticsSelectivityScale1999}
\bibinfo{author}{\bibfnamefont{G.~P.} \bibnamefont{van~der Laan}},
  \emph{\bibinfo{title}{Kinetics, Selectivity and Scale up of the
  {{Fischer-Tropsch}} Synthesis}} (\bibinfo{publisher}{Publisher not
  identified}, \bibinfo{address}{Netherlands?}, \bibinfo{year}{1999}), ISBN
  \bibinfo{isbn}{978-90-367-1011-4}.

\bibitem[{\citenamefont{Todic et~al.}(2016)\citenamefont{Todic, Nowicki,
  Nikacevic, and Bukur}}]{todicFischerTropschSynthesis2016}
\bibinfo{author}{\bibfnamefont{B.}~\bibnamefont{Todic}},
  \bibinfo{author}{\bibfnamefont{L.}~\bibnamefont{Nowicki}},
  \bibinfo{author}{\bibfnamefont{N.}~\bibnamefont{Nikacevic}},
  \bibnamefont{and} \bibinfo{author}{\bibfnamefont{D.~B.} \bibnamefont{Bukur}},
  \bibinfo{journal}{Catalysis Today} \textbf{\bibinfo{volume}{261}},
  \bibinfo{pages}{28} (\bibinfo{year}{2016}), ISSN \bibinfo{issn}{09205861}.

\bibitem[{\citenamefont{Vervloet et~al.}(2012)\citenamefont{Vervloet, Kapteijn,
  Nijenhuis, and Van~Ommen}}]{vervloetFischerTropschReaction2012}
\bibinfo{author}{\bibfnamefont{D.}~\bibnamefont{Vervloet}},
  \bibinfo{author}{\bibfnamefont{F.}~\bibnamefont{Kapteijn}},
  \bibinfo{author}{\bibfnamefont{J.}~\bibnamefont{Nijenhuis}},
  \bibnamefont{and} \bibinfo{author}{\bibfnamefont{J.~R.}
  \bibnamefont{Van~Ommen}}, \bibinfo{journal}{Catalysis Science \& Technology}
  \textbf{\bibinfo{volume}{2}}, \bibinfo{pages}{1221} (\bibinfo{year}{2012}),
  ISSN \bibinfo{issn}{2044-4753, 2044-4761}.

\bibitem[{\citenamefont{Davis and
  Occelli}(2010)}]{davisAdvancesFischerTropschSynthesis2010}
\bibinfo{editor}{\bibfnamefont{B.~H.} \bibnamefont{Davis}} \bibnamefont{and}
  \bibinfo{editor}{\bibfnamefont{M.~L.} \bibnamefont{Occelli}}, eds.,
  \emph{\bibinfo{title}{Advances in {{Fischer-Tropsch}} Synthesis, Catalysts,
  and Catalysis}}, no. \bibinfo{number}{128} in \bibinfo{series}{Chemical
  Industries} (\bibinfo{publisher}{CRC Press}, \bibinfo{address}{Boca Raton},
  \bibinfo{year}{2010}), ISBN \bibinfo{isbn}{978-1-4200-6256-4}.

\bibitem[{\citenamefont{Virtanen et~al.}(2020)\citenamefont{Virtanen, Gommers,
  Oliphant, Haberland, Reddy, Cournapeau, Burovski, Peterson, Weckesser, Bright
  et~al.}}]{Virtanen2020}
\bibinfo{author}{\bibfnamefont{P.}~\bibnamefont{Virtanen}},
  \bibinfo{author}{\bibfnamefont{R.}~\bibnamefont{Gommers}},
  \bibinfo{author}{\bibfnamefont{T.~E.} \bibnamefont{Oliphant}},
  \bibinfo{author}{\bibfnamefont{M.}~\bibnamefont{Haberland}},
  \bibinfo{author}{\bibfnamefont{T.}~\bibnamefont{Reddy}},
  \bibinfo{author}{\bibfnamefont{D.}~\bibnamefont{Cournapeau}},
  \bibinfo{author}{\bibfnamefont{E.}~\bibnamefont{Burovski}},
  \bibinfo{author}{\bibfnamefont{P.}~\bibnamefont{Peterson}},
  \bibinfo{author}{\bibfnamefont{W.}~\bibnamefont{Weckesser}},
  \bibinfo{author}{\bibfnamefont{J.}~\bibnamefont{Bright}},
  \bibnamefont{et~al.}, \bibinfo{journal}{Nature Methods}
  \textbf{\bibinfo{volume}{17}}, \bibinfo{pages}{261–272}
  (\bibinfo{year}{2020}), ISSN \bibinfo{issn}{1548-7105},
  \urlprefix\url{http://dx.doi.org/10.1038/s41592-019-0686-2}.

\bibitem[{\citenamefont{Baydin et~al.}(2018)\citenamefont{Baydin, Pearlmutter,
  Radul, and Siskind}}]{JMLR:v18:17-468}
\bibinfo{author}{\bibfnamefont{A.~G.} \bibnamefont{Baydin}},
  \bibinfo{author}{\bibfnamefont{B.~A.} \bibnamefont{Pearlmutter}},
  \bibinfo{author}{\bibfnamefont{A.~A.} \bibnamefont{Radul}}, \bibnamefont{and}
  \bibinfo{author}{\bibfnamefont{J.~M.} \bibnamefont{Siskind}},
  \bibinfo{journal}{Journal of Machine Learning Research}
  \textbf{\bibinfo{volume}{18}}, \bibinfo{pages}{1} (\bibinfo{year}{2018}),
  \urlprefix\url{http://jmlr.org/papers/v18/17-468.html}.

\bibitem[{\citenamefont{Krantz and Parks}(2013)}]{Krantz2013}
\bibinfo{author}{\bibfnamefont{S.~G.} \bibnamefont{Krantz}} \bibnamefont{and}
  \bibinfo{author}{\bibfnamefont{H.~R.} \bibnamefont{Parks}},
  \emph{\bibinfo{title}{The Implicit Function Theorem: History, Theory, and
  Applications}} (\bibinfo{publisher}{Springer New York},
  \bibinfo{year}{2013}), ISBN \bibinfo{isbn}{9781461459811},
  \urlprefix\url{http://dx.doi.org/10.1007/978-1-4614-5981-1}.

\bibitem[{\citenamefont{Smith et~al.}(2005)\citenamefont{Smith, Van~Ness, and
  Abbott}}]{smith2005introduction}
\bibinfo{author}{\bibfnamefont{J.}~\bibnamefont{Smith}},
  \bibinfo{author}{\bibfnamefont{H.}~\bibnamefont{Van~Ness}}, \bibnamefont{and}
  \bibinfo{author}{\bibfnamefont{M.}~\bibnamefont{Abbott}},
  \emph{\bibinfo{title}{Introduction to Chemical Engineering Thermodynamics}},
  CHEMICAL ENGINEERING SERIES (\bibinfo{publisher}{McGraw-Hill Education},
  \bibinfo{year}{2005}), ISBN \bibinfo{isbn}{9780073104454},
  \urlprefix\url{https://books.google.com.ua/books?id=c7J4TRxbGq8C}.

\bibitem[{\citenamefont{Marsden and Tromba}(2003)}]{Marsden2003-nz}
\bibinfo{author}{\bibfnamefont{J.}~\bibnamefont{Marsden}} \bibnamefont{and}
  \bibinfo{author}{\bibfnamefont{A.}~\bibnamefont{Tromba}},
  \emph{\bibinfo{title}{Vector Calculus}} (\bibinfo{publisher}{W.H. Freeman},
  \bibinfo{address}{New York, NY}, \bibinfo{year}{2003}),
  \bibinfo{edition}{5th} ed.

\bibitem[{\citenamefont{Rudin}(1976)}]{Rudin1976-nn}
\bibinfo{author}{\bibfnamefont{W.}~\bibnamefont{Rudin}},
  \emph{\bibinfo{title}{Principles of mathematical analysis}}, International
  series in pure and applied mathematics (\bibinfo{publisher}{McGraw-Hill
  Professional}, \bibinfo{address}{New York, NY}, \bibinfo{year}{1976}),
  \bibinfo{edition}{3rd} ed.

\bibitem[{\citenamefont{Kelley}(2003)}]{Kelley2003-ob}
\bibinfo{author}{\bibfnamefont{C.~T.} \bibnamefont{Kelley}},
  \emph{\bibinfo{title}{Solving nonlinear equations with Newton's method}},
  Fundamentals of algorithms (\bibinfo{publisher}{Society for Industrial and
  Applied Mathematics}, \bibinfo{address}{Philadelphia, Pa.},
  \bibinfo{year}{2003}).

\bibitem[{\citenamefont{Wang et~al.}(2001)\citenamefont{Wang, Xu, Xiang, Li,
  and Zhang}}]{Wang2001}
\bibinfo{author}{\bibfnamefont{Y.-N.} \bibnamefont{Wang}},
  \bibinfo{author}{\bibfnamefont{Y.-Y.} \bibnamefont{Xu}},
  \bibinfo{author}{\bibfnamefont{H.-W.} \bibnamefont{Xiang}},
  \bibinfo{author}{\bibfnamefont{Y.-W.} \bibnamefont{Li}}, \bibnamefont{and}
  \bibinfo{author}{\bibfnamefont{B.-J.} \bibnamefont{Zhang}},
  \bibinfo{journal}{Industrial and Engineering Chemistry Research}
  \textbf{\bibinfo{volume}{40}}, \bibinfo{pages}{4324} (\bibinfo{year}{2001}),
  ISSN \bibinfo{issn}{1520-5045},
  \urlprefix\url{http://dx.doi.org/10.1021/ie010080v}.

\end{thebibliography}
 
\newpage
\section*{Appendix: A}

\renewcommand\thefigure{A\arabic{figure}}    
\renewcommand\thetable{A\arabic{table}}    
\setcounter{figure}{0}
\setcounter{table}{0}

\begin{figure}[h!]
	\centering
	\subfigure[]{\includegraphics[width=0.27\textwidth]{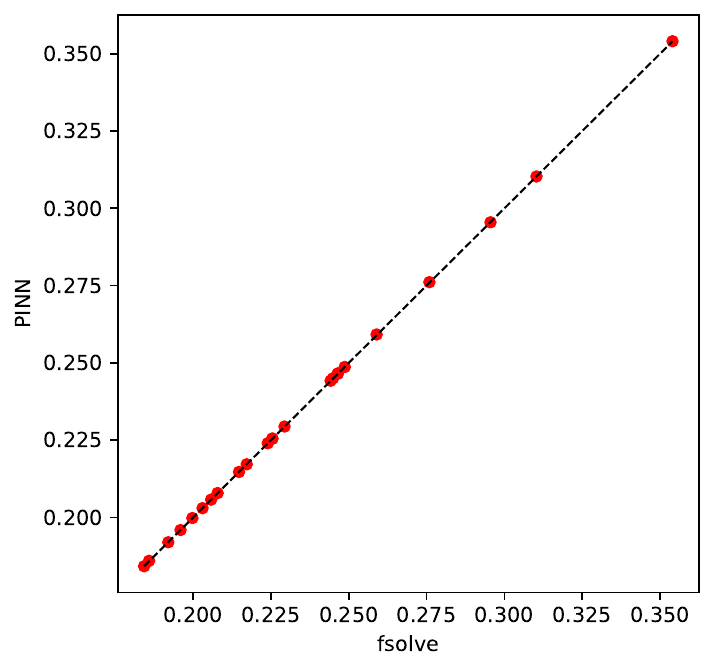}} 
	\subfigure[]{\includegraphics[width=0.32\textwidth]{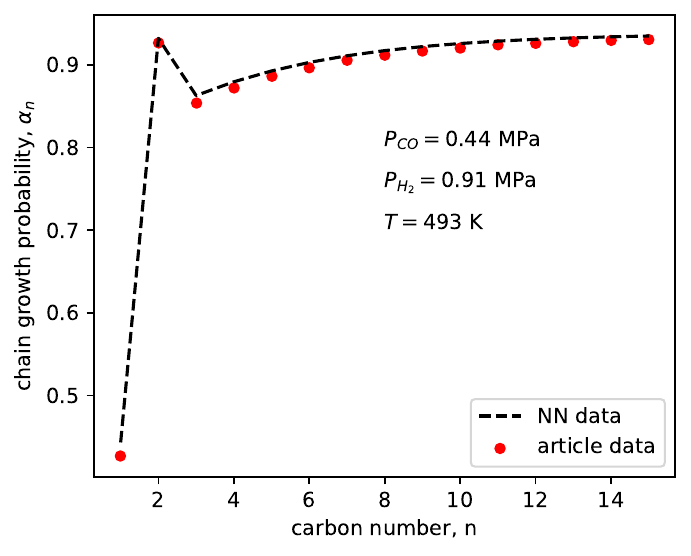}} 
	\subfigure[]{\includegraphics[width=0.32\textwidth]{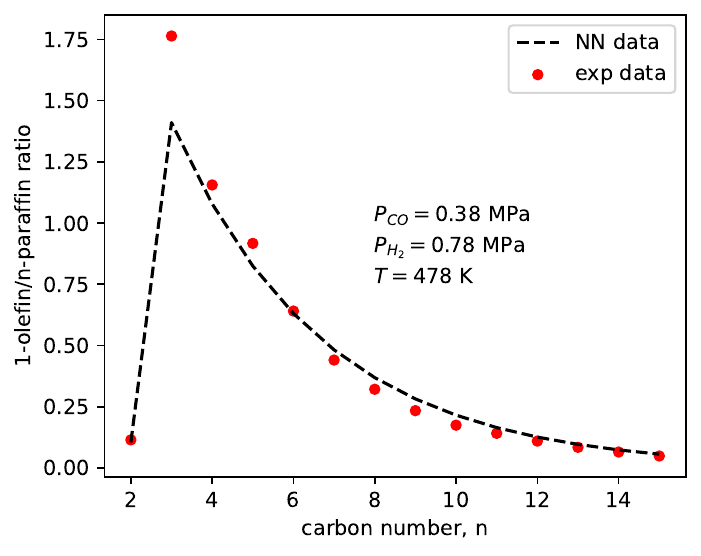}}
	\caption{(a) PINN-derived vs. \textit{fsolve}-derived values of $ \tilde{S} $ using the Experimental Dataset (b) Dependence of chain growth probability on carbon number for one of the experimental points calculated by proposed method and presented in [\cite{todicCOinsertionMechanismBased2014}] (c) Comparison of the 1-olefin/paraffin reaction rate ratios calculated experimentally and by the proposed method for one of the experimental point}
	\label{fig:m33}
\end{figure}

The practical applicability of the proposed approach was assessed using the experimental dataset. It shows the results for one practical reactor use, where the catalyst remains the same. In this dataset, the lumped parameters were calculated from 24 experimental data points provided in [\cite{todicCOinsertionMechanismBased2014}].
The $ \tilde{c}_0 $ values were within 0.164--0.217, $ \tilde{c}_S $ within 0.0008--0.009, $ \xi $ within 0.055--0.057, $ A $ within 0.029--0.079, and $ f $ within 0.758--0.769. Fig.~\ref{fig:m33}(a) compares $\tilde{S}$ values obtained by \textit{fsolve} with those predicted by the trained NN, demonstrating good agreement across all 24 experimental data points. The mean relative error of the NN predictions was $0.0237\%$, with a maximum relative error of $0.0928\%$. 

Chain growth probabilities, $\alpha_j$, were calculated for one experimental point. Fig.~\ref{fig:m33}(b) shows the dependence of $\alpha_j$ on the carbon number $j$, obtained using NN-derived $d\tilde{S}$ values, compared to the corresponding function presented in [\cite{todicCOinsertionMechanismBased2014}]. The results align closely, indicating the suitability of the NN approach for chain growth probability calculations.

These $\alpha_j$ values were further used to calculate reaction rates for paraffins and 1-olefins. Fig.~\ref{fig:m33}(c) illustrates the 1-olefin/paraffin reaction rate ratio as a function of carbon number for one experimental point, as calculated by trained NN and presented in [\cite{todicCOinsertionMechanismBased2014}] for one experimental point. The NN-derived results closely replicate the curve's shape and agree with most data points.
A notable discrepancy for $\alpha_3$ may be attributed to a reported error of $23.5\%$ in the established values of $K_i$ [\cite{todicCOinsertionMechanismBased2014}].

\end{document}